\documentclass[10pt,fleqn]{article}
\usepackage[nolist]{acronym}
\usepackage[fleqn]{mathtools}
\usepackage{cite}
\usepackage{dsfont}
\usepackage{enumitem}
\usepackage{mathrsfs}
\usepackage{pifont}
\usepackage{newpxtext}
\usepackage{newpxmath}
\usepackage{scalerel}
\usepackage{xspace}
\usepackage{microtype}
\usepackage[margin=1in]{geometry}
\usepackage[linkcolor=blue,urlcolor=blue,citecolor=magenta,colorlinks=true]{hyperref}
\usepackage[noabbrev]{cleveref}
\newcommand{\ii}{\mathop{}\!\mathrm{i}\!\mathop{}}
\newcommand{\ee}{\mathrm{e}}
\newcommand{\rep}[1]{\ensuremath{{\boldsymbol{#1}}}}

\DeclareMathOperator{\re}{Re}
\DeclareMathOperator{\im}{Im}
\makeatletter
\g@addto@macro\bfseries{\boldmath}
\newcommand*{\defeq}{\mathchoice{\mathrel{\rlap{%
\raisebox{0.24ex}{$\m@th\cdot$}}%
\raisebox{-0.24ex}{$\m@th\cdot$}}%
=}{\mathrel{\rlap{%
\raisebox{0.24ex}{$\m@th\cdot$}}%
\raisebox{-0.24ex}{$\m@th\cdot$}}%
=}{\mathrel{\rlap{%
\raisebox{0.08ex}{\small$\m@th\cdot$}}%
\raisebox{-0.28ex}{\small$\m@th\cdot$}}%
=}{\mathrel{\rlap{%
\raisebox{0.08ex}{\tiny$\m@th\cdot$}}%
\raisebox{-0.28ex}{\tiny$\m@th\cdot$}}%
=}}
\newcommand*{\eqdef}{\mathchoice{=\mathrel{\rlap{%
\raisebox{0.24ex}{$\m@th\cdot$}}%
\raisebox{-0.24ex}{$\m@th\cdot$}}}{%
=\mathrel{\rlap{%
\raisebox{0.24ex}{$\m@th\cdot$}}%
\raisebox{-0.24ex}{$\m@th\cdot$}}}{%
=\mathrel{\rlap{%
\raisebox{0.08ex}{\small$\m@th\cdot$}}%
\raisebox{-0.28ex}{\small$\m@th\cdot$}}}{%
=\mathrel{\rlap{%
\raisebox{0.08ex}{\tiny$\m@th\cdot$}}%
\raisebox{-0.28ex}{\tiny$\m@th\cdot$}}}%
}
\newcommand*{\transpose}{
    {\mathpalette\@transpose{}}%
    }
\newcommand*{\@transpose}[2]{%
    \raisebox{\depth}{$\m@th#1\intercal$}%
}
\makeatother
\newcommand{\RequirementSymbol}[2][]{\scalerel*{\ensuremath{\vcenter{\hbox{\includegraphics{r#2.pdf}}}}}{\ensuremath{M^1_2}}}
\newcommand{\Requirement}[2][]{\ifcase#2\relax\or 
\hyperref[item:modular_invariance]{\RequirementSymbol{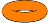}}\xspace
\or 
\hyperref[item:depend_on_tau_only]{\RequirementSymbol{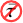}}\xspace
\or 
\hyperref[item:finite]{\RequirementSymbol{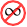}}\xspace
\fi}%
\newlist{requirements}{enumerate}{1}
\setlist[requirements]{label=(\Roman*),ref=(\Roman*),leftmargin=*, widest=abcd}
\crefname{requirementsi}{requirement}{requirements}
\Crefname{requirementsi}{Requirement}{Requirements}

\begin{document}
\title{Aspects of Modular Flavor Symmetries\footnotemark\footnotetext{Based on invited talks at the 2024 Moriond EW and QCD sessions, to appear in the proceedings.}}

\date{}
\author{Michael Ratz\\*[20pt]
\begin{minipage}{\linewidth}
\begin{center}
              {\itshape\small Department of Physics and Astronomy, University of California, Irvine, CA~92697-4575, USA}
\end{center}
\end{minipage}}

\maketitle
\begin{abstract}
Modular flavor symmetries refers to scenarios in which fermion masses respect modular symmetries. 
Such scenarios have been studied in the bottom-up approach and have an explicit realization in string theory. 
They rely on the remarkable properties of vector-valued modular forms.
\end{abstract}

\section{Introduction}

The \ac{SM} is remarkably successful in summarizing our current understanding of particle physics.
Currently experiments do not give us unambiguous hints for how particle physics beyond the \ac{SM} may look like.
At the same time, the \ac{SM} does not provide us with a theory of flavor, i.e.\ rather than explaining the patterns of fermion masses it only describes them by introducing more than 20 continuous parameters.

\section{Theories of flavor}

Developing theories of flavor has been a longstanding theme in beyond the \ac{SM} physics. 
Two main schemes emerged:
\begin{enumerate} 
 \item mass hierarchies from \ac{FN} and \ac{RS} models, 
 \item flavor structures from non-Abelian discrete flavor symmetries~\cite{Kaplan:1993ej}.  
\end{enumerate}
While the former can convincingly explain hierarchies, they typically introduce numerous extra parameters e.g.\ in the coefficients of terms in the Lagrange density. 
On the other hand, non-Abelian discrete symmetries allow one to avoid relative coefficients. 
For instance, consider the alternating group of order 4, $A_4$. 
At leading order the neutrino mass matrix reads~\cite{Altarelli:2005yp}
\begin{equation}\label{eq:m_nu_A_4_Traditional}
 m_\nu=\frac{v_u^2}{\Lambda}\begin{pmatrix}
  2v_1 & -v_3 & -v_2 \\
  -v_3 & 2v_2 & -v_1 \\
  -v_2 & -v_1 & 2v_3 
 \end{pmatrix}\;,
\end{equation}
where the $v_i$ denote the components of the \ac{VEV} of a triplet flavon, and $v_u$ stands for the \ac{VEV} of the $u$-type Higgs of the \ac{MSSM}. 
This means that the structure of the neutrino mass matrix is fixed by the $A_4$ symmetry.
However, it is nontrivial to explain \ac{VEV} patterns in which all $v_i$ are nonvanishing and their ratios are hierarchical. 
This conundrum sometimes gets referred to as \ac{VEV} alignment problem (see e.g.\ \cite{Holthausen:2011vd} for a discussion). 

\begin{figure}[htb]
 \centering\includegraphics[width=0.95\linewidth]{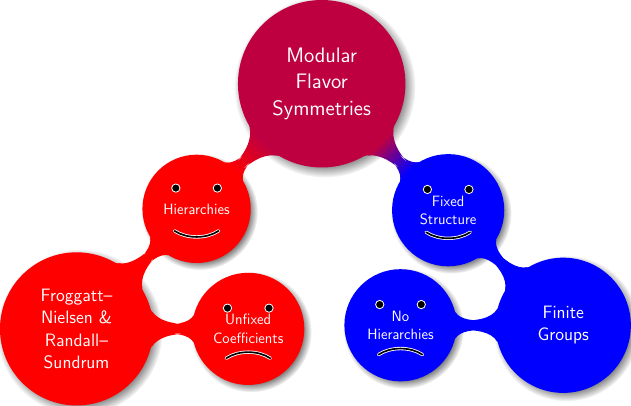}
 \caption{Modular flavor symmetries give rise to the attractive features of \ac{FN} and \ac{RS} models as well as non-Abelian flavor symmetries while avoiding their problematic aspects.}   
 \label{fig:ModularMindmap}    
\end{figure}

More recently, a new scheme emerged: modular flavor symmetries~\cite{Feruglio:2017spp}.\footnote{It is beyond the scope, and page limit, of these proceedings to fully summarize all related activities, see e.g.~\cite{Feruglio:2019ybq,Almumin:2022rml,Kobayashi:2023zzc,Ding:2023htn,Ding:2024ozt} for reviews and more references.} 
While the underlying symmetries have infinitely many elements, this approach hosts non-Abelian finite groups and naturally gives rise to hierarchies (cf.\ \Cref{fig:ModularMindmap}). 

\section{Modular invariance and flavor}
\label{sec:ModularInvariance}

Modular symmetries are, in a way, the symmetries of tori. 
A torus defines a lattice, and modular transformations are transformations of the lattice basis vectors which are such that the new basis vectors span the same lattice. 
These transformations form the infinite group 
\begin{equation}
 \mathrm{SL}(2,\mathds{Z})
 =
 \left\{\gamma=\begin{pmatrix}
  a & b\\ c & d     
 \end{pmatrix}\;;~a,b,c,d\in\mathds{Z}~\wedge~a\,d-b\,c=1\right\}\;.
\end{equation}
The so-called half-period ratio $\tau$ transforms as 
\begin{equation}\label{eq:ModularTrafoTau}
 \tau\xmapsto{~\gamma~}\frac{a\,\tau+b}{c\,\tau+d}\;,      
\end{equation}
and `sees' only the group $\Gamma\defeq\mathrm{SL}(2,\mathds{Z})/\mathds{Z}_2$, where the nontrivial $\mathds{Z}_2$ element is given by $\operatorname{diag}(-1,-1)$.
Models of flavor are often based on the so-called congruence subgroups 
\begin{equation}
       \Gamma(N)
       =
       \left\{\gamma\in\Gamma\;;\;
       \begin{pmatrix}
        a & b\\ c & d     
       \end{pmatrix}=\begin{pmatrix}
        1 & 0 \\ 0 & 1      
       \end{pmatrix}\pmod N\right\}\;.             
\end{equation} 
The quotients $\Gamma_N\defeq\Gamma/\Gamma(N)$ are finite groups, for instance $\Gamma_3\simeq A_4$. 
The important ingredient to tackle the flavor problem are the so-called \ac{VVMF}, which, by definition, transform under \eqref{eq:ModularTrafoTau} as 
\begin{equation}\label{eq:TrafoVectorValuedModularForm}
       f_i(\gamma\,\tau)=(c\tau+d)^{k}\,
       \left[\rho_n(\gamma)\right]_{ij}\, f_j(\tau)\;.      
\end{equation}
Here, the $f_i$ furnish an $n$-dimensional representation under $\Gamma_N$, $\rho_n(\gamma)$ is a representation matrix, and $k$ denotes the modular weight. 
The $f_i$ are further holomorphic functions of $\tau$ and do not possess poles. 
Crucially, these requirements make the \ac{VVMF} unique. 

This can be used to construct theories of flavor in which Yukawa couplings and mass matrices are modular forms. 
For instance, in the Feruglio model~\cite{Feruglio:2017spp} the pendant of \eqref{eq:m_nu_A_4_Traditional}, i.e.\ neutrino mass matrix, is proportional to
\begin{equation}\label{eq:m_nu}
m^{(\nu)}\propto\mathcal{M}_\nu(\tau)=\begin{pmatrix}
  2Y_1(\tau) & -Y_3(\tau) & -Y_2(\tau) \\
  -Y_3(\tau) & 2Y_2(\tau) & -Y_1(\tau) \\
  -Y_2(\tau) & -Y_1(\tau) & 
  2Y_3(\tau) 
 \end{pmatrix}\;.
\end{equation}
The proportionality factors contain the \ac{VEV} of the so-called $u$-type Higgs of the \ac{MSSM}, the see-saw scale $\Lambda$ as well as some factors from the K\"ahler metric which depend on $\tau$ and $\bar\tau$.
Crucially, the $Y_i$ are known modular functions, which can be written as~\cite{Liu:2021gwa}
\begin{equation}\label{eq:Y}
       \begin{pmatrix}
       Y_1(\tau)\\ Y_2(\tau)\\ Y_3(\tau) 
      \end{pmatrix}=\begin{pmatrix}
        \bigl(X_2(\tau)\bigr)^2 \\  \sqrt{2}X_1(\tau)\, X_2(\tau) \\ -\bigl(X_1(\tau)\bigr)^2
        \end{pmatrix}
        \quad\text{where }\left\{\begin{aligned}
        X_1(\tau)&\defeq 3\sqrt{2}\frac{\eta^3(3\tau)}{\eta(\tau)}\;,\\
       X_2(\tau)&\defeq -3\frac{\eta^3(3\tau)}{\eta(\tau)}-\frac{\eta^3(\tau/3)}{\eta(\tau)}\;.
      \end{aligned}\right.
\end{equation}
Here, $\eta$ denotes the Dedekind $\eta$-function. 
For largish $t\defeq\im\tau$,
\begin{align}
 |X_1(\ii t)|&\xrightarrow{t\text{ large}}|X_1(\ii)|\,\ee^{-2\pi(t-1)/3}\simeq 0.5234\,\ee^{-2\pi(t-1)/3}\;,\\
 |X_2(\ii t)|&\xrightarrow{t\text{ large}}1\;.        
\end{align}
This reveals that the entries of $Y$ in \eqref{eq:Y} can accommodate hierarchies. 
Somewhat ironically, though, it turns out that the best-fit point of this model is at $\tau\simeq\ii$.\footnote{Generally, modular symmetries give rise to linearly realized symmetries at certain critical points~\cite{Feruglio:2022koo,Feruglio:2023mii,Ding:2024xhz}, which lead to characteristic patterns of mass matrices. Small departures of $\tau$ from the critical values can be related to \ac{SUSY} breaking~\cite{Knapp-Perez:2023nty}.}
In this model, 3 parameters predict 9 observables,
\begin{equation}
\left.\begin{array}{l}
\Lambda\\ \re\tau\\ \im\tau
\end{array}\right\}\xrightarrow{\text{predict}}
\left\{\begin{array}{l}
\text{3 mass eigenvalues }m_i\;,\\ \text{3 mixing angles }\theta_{ij}\;,\\ 
\text{3 phases (1 Dirac \& 2 Majorana)\;.}
\end{array}\right.
\end{equation}
This means that this model is dramatically overconstrained. 
It is still remarkably consistent with data, and can be made fully consistent by introducing one, admittedly ad hoc, parameter \cite{Criado:2018thu}.

\section{Modular invariant holomorphic observables}

The predictive power of modular flavor symmetries relies on three key ingredients,
\begin{requirements}
       \item[\RequirementSymbol{1}] modular covariance/invariance (cf.\ \Cref{sec:ModularInvariance}),\label{item:modular_invariance}  
       \item[\RequirementSymbol{2}] meromorphy, i.e.\ the couplings do not depend on $\bar\tau$, and\label{item:depend_on_tau_only} 
       \item[\RequirementSymbol{3}] the couplings remain finite for all values of $\tau$.\label{item:finite}
\end{requirements}
In the respective models, these requirements fix the superpotential. 
It turns out that, in specific cases, they directly fix observables \cite{Chen:2024otk}. 
This is true e.g.\ in the Feruglio model~\cite{Feruglio:2017spp}, where \cite{Chen:2024otk}
\begin{equation}
 I_{ij}(\tau)\defeq 
 \frac{m_{ii}^{(\nu)} (\tau, \bar\tau) \,m_{jj}^{(\nu)} (\tau, \bar\tau)}{\bigl(m_{ij}^{(\nu)} (\tau, \bar\tau)\bigr)^2}
 =
 \frac{\bigl(\mathcal{M}_\nu(\tau)\bigr)_{ii}\,\bigl(\mathcal{M}_\nu(\tau)\bigr)_{jj}}{\bigl(\mathcal{M}_\nu(\tau)\bigr)_{ij}^2}  
\end{equation}
with $m^{(\nu)}$ from \eqref{eq:m_nu} are meromorphic modular invariant functions, i.e.\ fulfill requirements \Requirement{1} and \Requirement{2}. 
The $I_{ij}$ are known to be \ac{RG} invariant~\cite{Chang:2002yr}.
At the same time, the $I_{ij}$ are functions that depend only on the physical neutrino masses, mixing angles and phases. 
Two combinations of the $I_{ij}$ even fulfill \Requirement{3}. 
This allows one to obtain a large amount of information of the model directly from the theory of modular forms. 

\section{Origin of modular flavor symmetries}

While modular flavor symmetries have been proposed in the bottom-up approach, Yukawa couplings are known to be modular forms in certain explicit string models \cite{Chun:1989se,Erler:1992gt,Quevedo:1996sv,GrootNibbelink:2017usl}, including the so-called toroidal orbifolds (see \cite{Ramos-Sanchez:2024keh} for a recent review).  
The latter are, as their name suggests, based on tori and, therefore, it is not too surprising that they exhibit modular symmetries. 
Crucially, this means that modular flavor symmetries emerge from a consistent scheme of quantum gravity.
More recently, these symmetries have been explored further, which has led to the so-called eclectic scheme~\cite{Baur:2019iai,Nilles:2020tdp,Nilles:2020gvu,Baur:2020jwc}, which unifies modular symmetries, traditional family symmetries, $R$ symmetries and outer automorphisms like the $\mathcal{CP}$ transformation under one umbrella. 
Modular flavor symmetries arise also from magnetized tori \cite{Cremades:2004wa,Kikuchi:2021ogn,Almumin:2021fbk}.

Given that modular flavor symmetries have a \ac{UV} completion, it is instructive to compare bottom-up and top-down models~\cite{Nilles:2024iqp}. 
Both approaches use \ac{VVMF} to describe fermion masses.
In explicit string models, the modular weights of quarks and leptons (and other matter fields) are small, indeed often fractional, whereas in phenomenological models they are sometimes taken to be rather large, thus allowing for several contractions to contribute to the fermion mass matrices. 
This introduces several freely adjusted parameters in the bottom-up approach, giving rise to models that reproduce data, something that this is much harder in the top-down approach.

\section{Some open questions and challenges}

\subsection{Problems with kinetic terms}
\label{sec:KineticTerms}

While modular flavor symmetries constrain the superpotential very strongly, the non-holomorphic K\"ahler potential is far less restricted. 
In fact, the \ac{EFT} expansion of the K\"ahler potential \cite{Chen:2019ewa}
\begin{align}
 K&=
 \alpha_0\,\left(-\ii\tau+\ii\bar\tau\right)^{-1}\,
 \left(\overline{L}\,L\right)_{\rep{1}}
 +
 \sum_{k=1}^7\alpha_k
 \left(-\ii\tau+\ii\bar\tau\right)\,\left(Y\,L\,\overline{Y}\,\overline{L}
 \right)_{\rep{1}, \, k}+\dots
\end{align}
contains infinitely many terms beyond the so-called minimal term proportional to $\alpha_0$.
Even moderate values of the $\alpha_{i>0}$ coefficients lead to changes of the predicted values of observables which exceed the experimental error bars by far \cite{Chen:2019ewa}. 
It should be stressed that higher-order terms in the K\"ahler potential are essential for the viability of the model as the soft \ac{SUSY} breaking terms rely on them. 
Furthermore, attempts to suppress them completely in \ac{UV} complete models appear hard, if not impossible \cite{Kachru:2007xp}.
Nonetheless, a proof-of-principle solution to the problem in which the impact of these terms does not exceed the experimental uncertainties exists \cite{Chen:2021prl}, but more research in this direction will be needed to arrive at a fully convincing picture. 
Note that these extra terms can also be used in order to improve the fit of a given model \cite{Baur:2022hma}, though at the expense of introducing more parameters and thus reducing the predictive power of the scheme. 

\subsection{Is \ac{SUSY} indispensable?}

A key ingredient of modular flavor models so far is holomorphy, \Requirement{2}, which is implemented by imposing \ac{SUSY}.
Given the absence of clear signals for low-energy \ac{SUSY}, one may ask whether it is possible to define a nonsupersymmetric version of modular flavor symmetries. 
It has been pointed out \cite{Cremades:2004wa,Almumin:2021fbk} that in certain scenarios low-energy \ac{SUSY} may not be crucial to have mass matrices described by modular forms. 
However, to date there is no complete model illustrating these points. 

\section{Outlook}

Modular flavor symmetries are a rather new and exciting scheme allowing us to tackle the flavor problem.
This scheme may constitute the most concrete way of figuring out what completes the \ac{SM} in the \ac{UV}.
Models in this realm can be highly predictive, and can be tested in the foreseeable future. 
It remains a challenge for the future to build predictive models which are fully consistent with observation and in which the theoretical uncertainties are smaller than the experimental ones. 

\subsection*{Acknowledgments}

It is a pleasure to thank 
the organizers of Moriond 2024 for putting together such great meetings, and 
Y.~Almumin, 
M.-C.~Chen, 
V.~Knapp-Perez,
X.~Li,
X.-G.~Liu,
O.~Medina,
H.P.~Nilles,
M.~Ramos-Hamud,
S.~Ramos-S\'anchez
and 
S.~Shukla 
for fruitful collaborations on this topic.
This work is supported by the National Science Foundation, under Grant No.\ PHY-1915005.

\begin{acronym}
       \acro{EFT}{effective field theory}
       \acro{FN}{Froggatt--Nielsen~\protect\cite{Froggatt:1978nt}}
       \acro{MSSM}{minimal supersymmetric standard model}
       \acro{RG}{renormalization group}
       \acro{RGE}{renormalization group equation}
       \acro{RS}{Randall--Sundrum~\protect\cite{Randall:1999ee}}
       \acro{SM}{standard model}
       \acro{SUSY}{supersymmetry}
       \acro{UV}{ultraviolet}
       \acro{VEV}{vacuum expectation value}
       \acro{VVMF}{vector-valued modular forms~\protect\cite{Liu:2021gwa}}
\end{acronym}

\bibliographystyle{utphys}
\bibliography{Orbifold}

\end{document}